\documentclass[fleqn,twoside]{article}%
\usepackage{amsmath}
\usepackage{amsfonts}
\usepackage{amssymb}
\usepackage{graphicx}
\usepackage{bm}

\newcommand{\be}{\begin{equation}}
\newcommand{\ee}{\end{equation}}
\newcommand{\bey}{\begin{eqnarray}}
\newcommand{\eey}{\end{eqnarray}}
\def\bes{\begin{equation}\begin{split}&}
\newcommand{\bi}{\bibitem}

\title  {Noether Symmetry of  Palatini $F(\Re)$ gravity.}
\author{Nayem Sk$^{1}$.\\
~~~~~~~~~~~~~~~\\
$^1$Dept. of Physics, University of Kalyani, West Bengal, India - 741235\\}
\begin{document}
\maketitle
\footnote{
Electronic address:\\
$^{1}$nayemsk1981@gmail.com\\}
\begin{abstract}
 In metric formalism, Noether symmetry of F(R) theory of gravity in vacuum and in the presence of pressureless dust yields $F(R)\propto R^\frac{3}{2}$ along with the conserved current $\frac{d}{dt} (a\sqrt R)$ in Robertson-Walker metric and nothing else. However, Roshan et.al. has claimed in $\mathrm{Phys. Lett. \textbf{B668}, 238 (2008)}$ \cite{o} that Noether symmetry in the context of Palatini $F(\Re)$ theory of gravity admits $F(\Re)\propto \Re^{s}$, (where $s$ is arbitrary) in matter domain era in Friedmann- Robertson-Walker universe. But, it has been shown that the conserved current obtained under the process does not satisfy the field equations in general. Here, it is shown that Noether Symmetry admits $F(\Re)\propto\Re^\frac{3}{2}$ along with a conserved current $\Sigma_{0} {a}[\frac{\dot a}{a}+ \frac{\dot\Phi}{2\Phi}]$ in Palatini gravity. Thus, their claim is not right.

\end{abstract}

\section{Introduction}
Modified theory of gravity has been advocated as a strong contender to alternative theory of dark energy, since it can possibly reconcile early inflation with late time cosmic acceleration \cite{no, cd}. However, neither setting a form of $F(R)$ by hand \cite{felice}, nor reconstruct it from the history of cosmic evolution \cite{nojiri05} is viable option since, arbitrarily many such models might pass the solar test and admit cosmological bound, and there would be no means to select the appropriate one. Rather, it should be chosen following some fundamental physical consideration, like loop quantum gravity or by invoking Noether symmetry. Loop quantum gravity does not provide a term suitable for late time cosmic acceleration. Therefore, Noether symmetry is preferred \cite{cap7, vak, sar2}. Noether symmetry had been applied initially by Rugeiro and his co-workers \cite{ritis1, ritis2} in cosmology, to find the form of the scalar potential in scalar-tensor theory of gravity. The potential so found was exponential, which is suitable to drive inflation in the early stage of cosmic evolution. In view of such exciting results, till date, there has been numerous attempts in this field. Of particular importance are its application in scalar-tensor theory of gravity \cite{ca, san,l} and in higher order theory of gravity \cite{cap7, vak, sar2, vil, cap1, cap2, cap3, cap4, cap5, cap6,  cap8,  sar1, higher1, higher2}.\\

\noindent
$F(R)$ theory of gravity, is a class of extended theories of gravity (ETGs) that represents one of the simplest modifications to GTR \cite{buc}. Such a theroy arises from replacing the Ricci scalar $R$ in the Einstein-Hilbert action, by a more general and arbitrary function of $F(R)$ \cite{felice, soi}. There exists three flavours of $F(R)$ theory of gravity, by the names, metric, Palatini, and metric affine formalisms. While different formalisms of the theory of general relativity lead to the same field equations due to the linearity of the action in the Ricci scalar $R$, the same is not expected for higher order theories of gravity \cite{lau}. Metric formalism contains fourth order field equations. An advantage of Palatini formulation rely on second order field equations which turns out to be more easy to solve. In this sense the Palatini formalism is more easy to handle and simpler to analyze then the corresponding metric formalism. A more extensive description may be found in \cite{soi1}. \\

\noindent
Results (metric formalism) in a nutshell are (i) Noether symmetry gives $F(R)=F_0 R^{\frac{3}{2}}$ along with a conserved current $\frac{d}{dt} (a\sqrt R)$ in R-W metric when coupled to pressure-less dust or in vacuum \cite{cap7,vak, sar2,l}. (ii) It does not yield any symmetry when the configuration space is enlarged by including scalar field or in anisotropic model \cite{sar1}. (iii) Inclusion of gauge term does not improve the situation \cite{i, j, k}. (iv) Exact solution in R-W metric does not satisfy all cosmological data and a linear term may be necessary in addition \cite{sar1}. (v) Metric variation of higher order theory of gravity requires to fix the Ricci scalar in addition to the metric tensor at the boundary. Fixing Ricci scalar at the boundary implies that the classical solutions are fixed once and forever to the de-Sitter or anti de-Sitter solutions. Noether Symmetry justify such requirement \cite{ruz,sar3}. (vi) A non-Noether conserved current exists in general for gravity when a non-minimally coupled scalar field is added to $F(R)$. It has been shown that it is useful to obtain improved exact solution \cite {sar1, k}. \\

\noindent
Here, Palatini $F(\Re)$ gravity is considered. In this formalism, the connection (usually torsion free), out of which the Ricci tensor is defined, is treated to be independent of the space-time metric, i.e., it is no longer the Levi-Civita connection. As a result, the action is varied both with respect to the metric tensor and the connection. In the case of Einstein-Hilbert action, Palatini variational techniques reproduces general theory of relativity (GTR) making the
connection to be the Levi-Civita one. The only difference is that the surface term does not arise in the process and thus the equivalence of the surface term with the black hole entropy is lost. However, for extended theory of gravity the two variational techniques yield different field equations and hence different Physics altogether. These issues have been thoroughly discussed in a recent review \cite{cd}. To cast the action containing $F(\Re)$ gravity in the canonical form in view of Palatini variational formalism it is required to first form a scalar-tensor equivalence of the theory. However, this theory has been shown to be equivalent to Brans–Dicke theory with $w = -\frac{3}{2}$ \cite{fla,olm}. \\

\noindent
Roshan et.al. has claimed \cite{o} that Noether symmetry in the context of Palatini $F(\Re)$ theory of gravity admits $F(\Re)\propto \Re^{s}$, (where $s$ is arbitrary) in matter domain era in Friedmann- Robertson-Walker universe. Here we  review the work and prove that such claim is not right. In the following, we first write down the expressions of action, field equations and Noether Symmetry for $F(\Re)$ Palatini gravity. Thereafter, Noether symmetry of the action which leads to explicit forms of $F(\Re)$ is also discussed. We then review the  Rosan et.al. work. Finally concluding remarks is presented in summary.\\

\section{Action, Field equations and Noether Symmetry}

In Palatini $F(\Re)$ gravity, one treats the metric and connection independently and varies the action with respect to each of them separately. The matter Lagrangian is assumed to be independent of the connection. The Palatini $F(\Re)$ action takes the form
\be S_{pal} =\frac{1}{2\kappa} \int  d^4 x\sqrt{-g}[ F(\Re)+ S_m(g_{\mu\nu},\Psi)],\ee
\noindent
where, $\kappa = 8\pi G $ and $F(\Re)$ is an arbitrary function of $\Re$ = $g^{\mu\nu}\Re_{\mu\nu}( \Gamma )$ and $\Gamma_{{\mu\nu}}^{\lambda}$ is the connection. The matter action $S_m$ depends on the matter fields $\Psi$ and $g^{\mu\nu}$. Varying equation (1) with respect to the metric $g^{\mu\nu}$ gives
\be F'(\Re)\Re-\frac{1}{2}F(\Re)g_{\mu\nu}=\kappa T_{\mu\nu},\ee
where the prime denotes differentiation with respect to $\Re$.
The trace of equation (2) gives us the structural equation of the spacetime controlling (2)
\be F'(\Re)\Re-2F(\Re)=\kappa T\ee
and now variation of equation (1) with respect to connection yields
\be \nabla_\lambda(\sqrt -g F'(\Re)g^{\mu\nu})=0 .\ee
In fact, taking into account how the Ricci tensor transforms under conformal transformations \cite {soi1, n}, it has been shown that
 \bes\label{p}\Re_{\mu\nu}=R_{\mu\nu}+\frac{3}{2[F'(\Re)]^2}(\nabla_\mu F'(\Re))(\nabla_\nu F'(\Re))\\&-\frac{1}{F'(\Re)}(\nabla_\mu\nabla_\nu-\frac{1}{2}g_{\mu\nu}\Box)F'(\Re),
 \end{split}\ee
contracting with $g^{\mu\nu}$ to obtain,
\be \Re = R+\frac{3}{2[F'(\Re)]^2}(\nabla_\mu F'(\Re))(\nabla_\nu F'(\Re))-\frac{3}{F'(\Re)}\Box F'(\Re),\ee
where $R$ is the Ricci scalar.

The matter action $S_m(g_{\mu\nu},\Psi)$ is
\be S_m(g_{\mu\nu},\Psi) = \frac{1}{2\kappa} \int  d^4 x\sqrt{-g}[ L_m  ],\ee
in which the Lagrangian $L_m $ corresponding to matter field which is denoted by  $ L_m = -\rho_{m0} a^3$. One usually introduce a new field \cite{soi1, n, a,a.1,a.2,a.3,a.4,a.5,a.6,a.7,a.8,a.9,a.10,a.11,a.12,a.13} $\chi=\Re$ by which the action (1) can be written as
\be S_{pal} =\frac{1}{2\kappa} \int  d^4 x\sqrt{-g}[ F(\chi)+ F'(\chi)(\Re-\chi)]+ S_m(g_{\mu\nu},\Psi),\ee
where prime denotes differentiation w.r.t $\Re$. Now variation with respect to $\chi$ leads to the equation
$F''(R)(\chi-\Re)=0$. If $F''(\Re)\ne0$, then $\chi=\Re$ inserting this result reproduces the action and then redefinition the field $\chi$ by $\Phi=F'(\Re)$ and setting
$U(\Phi)= \chi(\Phi)\Phi-F(\chi(\Phi))$ the action (8) takes the form
\be S_{pal} = \frac{1}{2\kappa} \int  d^4 x\sqrt{-g}[ \Phi \Re-U(\Phi)]+S_m(g_{\mu\nu},\Psi).\ee
\noindent

\be S_{pal} = \frac{1}{2\kappa} \int  d^4 x\sqrt{-g}[ \Phi R +\frac{3}{2\Phi}\partial_\mu\Phi\partial^\mu\Phi-U(\Phi)]+S_m(g_{\mu\nu},\Psi).\ee
\noindent
Using the F-R-W metric, the Scalar curvature takes the form \be R = - 6\left(\frac{\ddot a}{a} + \frac{ \dot a^2}{a^2}+ \frac{k}{a^2}\right). \ee
Equation (10) is the Brans-Dicke action with a potential $U(\Phi)$ and a Brans-Dicke parameter $\omega=-\frac{3}{2}$. Therefore there is a dynamical equivalence between $F(\Re)$ theory and a class of Brans-Dicke theories  with a potential. The important point in the said transformation is that the matter sector is remain unchanged. In particular, in this representation of $F(\Re)$ theories the matter action $S_m$ is independent of the scalar field $\Phi$.
In order to apply the Noether symmetry approach, in a FRW manifold the Lagrangian related to the action (10) takes the point-like form

 \bes\label{p1} L(a,\Phi,\dot a,\dot\Phi) = 6a^2 \dot a \dot \Phi + 6a \dot a^2\Phi-6k\Phi a + \frac{3a^3 \dot\Phi^2}{2\Phi} - a^3U(\Phi)\\& + 2\kappa \rho_{m0}
\end{split}\ee
The equation of motion can be obtained by varying the Lagrangian to $a$ and $\Phi$ , respectively as follows
\be \left[2\frac{\ddot a}{a} + \frac{\dot a^2}{a^2} + \frac{\ddot \Phi}{\Phi}+ 2 \frac{\dot a\dot \Phi}{a \Phi} + \frac{k}{a^2}-\frac{3 \dot\Phi^2}{4\Phi^2} + \frac{U}{2\Phi}\right] =0.\ee
\be \left[\frac{\ddot a}{a} + \frac{\dot a^2}{a^2} + \frac{\ddot \Phi}{2\Phi}+ \frac{3\dot a\dot \Phi}{2a \Phi} + \frac{k}{a^2}-\frac{ \dot\Phi^2}{4\Phi^2} + \frac{U_{,\Phi}}{6}\right] =0.\ee
Hamiltonian equation (H) is
\be \left[\frac{\dot a^2}{a^2} + \frac{\dot a\dot \Phi}{a \Phi} + \frac{k}{a^2}+\frac{ \dot\Phi^2}{4\Phi^2} + \frac{U}{6\Phi}-\frac{\kappa \rho_{m0}}{3a^3 \Phi}\right] =0.\ee
Noether theorem state that, if there exists a vector field X, for which the Lie derivative of a given Lagrangian L vanishes i.e. $\pounds_X L =X L = 0 $, the Lagrangian admits a Noether Symmetry and thus yields a conserved current. In the Lagrangian under consideration the configuration space is $M(a,\Phi)$ and the corresponding tangent space is $TM(a,\Phi,\dot a, \dot\Phi)$. Hence the generic infinitesimal generator of the Noether Symmetry is

 \be X = \alpha \frac{\partial }{\partial a}+\beta\frac{\partial }{\partial \Phi} +\dot\alpha \frac{\partial }{\partial\dot a}+ \dot\beta\frac{\partial }{\partial\dot \Phi} \ee
  The constant of motion is given by
  \be \Sigma = \alpha \frac{\partial L }{\partial\dot a}+ \beta\frac{\partial L }{\partial\dot \Phi} \ee
  The existence condition for symmetry of Noether Symmetry  $\pounds_X L= X L = 0 $, then leads to the following system of partial differential equations

  \begin{eqnarray}
   \alpha+\frac{a\beta}{\Phi} +2a\alpha_{,a}+\frac{a^2\beta_{,a}}{\Phi}  &=& 0\\
   3\alpha\Phi - a \beta +2a\Phi\beta_{,\Phi}+4\Phi^2\alpha_{,\Phi}&=& 0  \\
   2\alpha+a\alpha_{,a}+2\Phi\alpha_{,\Phi}+a\beta_{,\Phi}+\frac{a^2\beta_{,a}}{2\Phi} &=& 0\\
 6k\Phi(\alpha+\frac{a\beta}{\Phi})+ 3a^2\alpha U +a^3\beta U_{,\Phi}  &=&0
     \end{eqnarray}

 The above set of partial differential equations admit the following set of solutions, for $k = +1, 0, -1$
\begin{eqnarray}
    \alpha =ca(a^2\Phi)^{-\frac{(m+1)}{(2m+1)}} \\
     \beta =c(2m-1)\Phi(a^2\Phi)^{-\frac{(m+1)}{(2m+1)}}\\
        U(\Phi) = U_{0} \Phi^\frac{-3}{(2m-1)}
     \end{eqnarray}
One can find the form of $F(\Re)$, in view of equation (24) as,
\be F(\Re)=g(m)\Re^\frac{3}{2m+2},\ee
and the expression of conserved current reads
\be \Sigma= 3c(2m+1)a(a^2\Phi)^\frac{(m+1)}{(2m+1)}[2a\dot a \Phi+ a^2\dot\Phi].\ee
It is shown that the above conserved current satisfies the field equations (13) to (15) only for $m = 0$ otherwise not. Now, equation (26) gives the expression of conserved current for $m =0 $ is
\be \Sigma= \Sigma_{0} a[\frac{\dot a}{a}+ \frac{\dot\Phi}{2\Phi}].\ee
It is interesting that the reduced form of $F(\Re)$ is
\be F(\Re)=F_{0}\Re^\frac{3}{2}.\ee
Here, it has been shown that Noether Symmetry selects  $F(\Re)\propto\Re^\frac{3}{2}$ along with a conserved current $\Sigma_{0} a[\frac{\dot a}{a}+ \frac{\dot\Phi}{2\Phi}]$ in Palatini gravity. Again, it is also clear that the conserved current and the form of Ricci scalar (in view of equation 6) are also different from metric formalism.\\

\noindent
Let us now find out the cosmological solution for $F(\Re)\propto F_{0}\Re^\frac{3}{2}$ gravity. It can be written from equation (3),
\be F'(\Re)\Re-2F(\Re)= - \frac{\kappa \rho_{m0}}{a^3}.\ee
This equation can be express as an equation of $\Re$ in term of the cosmic scale factor, that is
\be \Re = \frac{2\kappa \rho_{m0}}{F_0 a^2}\ee
 Again, it is known that $\Phi=F'(\Re)$.
 Now the expression of $\Phi$ in term of the cosmic scale factor is
 \be \Phi = \frac{\Phi_0}{a},\ee
 where $\Phi_0 = 3 \sqrt{ \frac{\kappa F_0 \rho_{m0}}{2}}$.
 Therefore, the solution of the cosmic scale factor from the conserved current equation (27) is
 \be a(t)= \frac{\Sigma}{3c}t + a_0 ,\ee where $a_0$ is a constant of integration.
 However, this type of solution is not bad, such a coasting solution although fits SnIa data perfectly in the matter domain era \cite{sar1}.  This type of solution was also found analytically earlier \cite{gams}.\\

\section{Reviewing Rosan et.al. work}
\noindent
It has been shown that Noether symmetry of $F(\Re)$ theory of palatini  gravity also does not admit anything other than $\Re^\frac{3}{2}$. But Roshan et al. has been claimed \cite{o} that Noether symmetry admits $F(\Re) \propto \Re^s$ in palatini  gravity, where $s$ is arbitrary. In the following, let us review the Rosan et al. work. Rosan et al. used the redefinition $\Phi=\varphi^2$ in equation (12) and now the form of point-like Lagrangian (according to their manuscript) is
\be L(a,\varphi,\dot a,\dot\varphi) = 12a^2 \dot a\varphi \dot \varphi + 6a \dot a^2\varphi^2+ 6a^3 \dot\varphi^2 - a^3U(\varphi)-2\kappa \rho_{m0} ,\ee
here,\be R = - 6\left(\frac{\ddot a}{a} + \frac{ \dot a^2}{a^2}\right). \ee
The equations of motion for $a$ and $\varphi$ are
\be \left[\frac{\ddot a}{a} + \frac{\dot a^2}{2a^2} + \frac{\ddot \varphi}{\varphi}+ 2 \frac{\dot a\dot\varphi}{a\varphi} -\frac{\dot\varphi^2}{2\varphi^2} + \frac{U}{4\varphi^2}\right] =0\ee
\be \left[\frac{\ddot a}{a} + \frac{\dot a^2}{a^2} + \frac{\ddot \varphi}{\varphi}+ \frac{2\dot a\dot \varphi}{a \varphi} -\frac{ \dot\varphi^2}{2\varphi^2} + \frac{U_{,\Phi}}{12\varphi}\right] =0\ee
Hamiltonian equation (H) is
\be \left[\frac{\dot a^2}{a^2} + \frac{2\dot a\dot \varphi}{a \varphi} +\frac{ \dot\varphi^2}{\varphi^2} + \frac{U}{6\varphi^2}+\frac{\kappa \rho_{m0}}{3a^3 \varphi^2}\right] =0.\ee
Noether equations for the above point Lagrangian are

\bes\label{p2}
3A+2\varphi A_{,\varphi} +2a B_{,\varphi}  = 0\\&
\varphi A +2aB+2a\varphi A_{,a}+2a^2 B_{,a}= 0  \\&
2\varphi A+aB+ \varphi^2 A_{,\varphi}+a\varphi A{,a}+a^2B_{,a}+ a\varphi B_{,\varphi} = 0\\&
3a^2 A U +a^3B U_{,\varphi} =0
\end{split}\end{equation}
\noindent
Rosan et.al. has found the following form of A, B and $U(\varphi)$ in their manuscript

\begin{eqnarray}
    A =\beta a^n \varphi^{n-1} \\
     B = -\frac{2n+1}{2n}\beta a^{n-1}\varphi^n\\
        U(\varphi) = \lambda \varphi^\frac{6n}{1+2n} = \lambda \Phi^\frac{3n}{1+2n}
     \end{eqnarray}
They have been solved the equation (41) and found the form of $F(\Re)$ are
\be F(\Re)=g(n)\Re^\frac{3n}{n-1}\ee
\be F(\Re)=\alpha\Re -\lambda\alpha^\frac{3n}{1+2n}\ee
The existence of Noether symmetry means that there exists a constant of motion. The constant of motion for Lagrangian (33) is
\be \Sigma=-\frac{6\beta}{n}a^{n+1}\varphi^n[\dot a\varphi+a \dot\varphi]. \ee
Conserved current is not an independent equation, but rather it is the first integral of certain combination of the field equations. Thus, it is essential to check if the conserved current obtained this process satisfy the field equations, which was not performed by Rosan et.al. It is shown that the above conserved current satisfies the field equations (35) to (37) only for $n = -1$, otherwise not. Therefore, equation (44) yields for the value of  $n = -1$ is
\be \Sigma=\Sigma_{0} [\dot a+\frac{a \dot\varphi}{\varphi}] = \Sigma_{0} a[\frac{\dot a}{a}+ \frac{\dot\Phi}{2\Phi}]. \ee

This conserved current is identical with equation (27). Now equation (42) takes the form for $n = -1$ is
\be F(\Re)=g\Re^\frac{3}{2}.\ee

\noindent
It is clear that Noether Symmetry does not admit the arbitrary power of $\Re$ in Palatini $F(\Re)$ gravity except $ F(\Re)\propto\Re^\frac{3}{2}$, which is found earlier in equation (28). On the other hand the equation (43) does not valid, since $F''(\Re)\ne0$, which is mentioned above. Thus their \cite{o} claim is not right.\\
\section{Concluding remarks}
\noindent
In summary, earlier attempts (metric formalism) in finding Noether symmetry for F(R) theory of gravity in the vacuum and matter dominated era yields $F(R)= R^\frac{3}{2}$ along with a conserved current $\frac{d}{dt} (a\sqrt R)$ in R-W metric \cite{cap7, vak, sar2,l}. Such a form is not suitable to explain presently available cosmological data \cite{sar2}. Search of a better form of $F(R)$ taking a scalar field minimally or non-minimally into account failed to produce symmetry \cite{sar1}. It is also found that Noether symmetry does not yield anything new in the presence of gauge \cite{i, j, k}. Here, it has been shown  that Noether Symmetry selects  $F(\Re)\propto\Re^\frac{3}{2}$ along with a conserved current $\Sigma_{0} {a}[\frac{\dot a}{a}+ \frac{\dot\Phi}{2\Phi}]$ in Palatini gravity. It is also clear that the form of conserved currents and Ricci scalar are also different in both formalism (metric and Palatini) for higher order gravity. However, Rosan et.al. \cite{o} claimed that $F(\Re)$ in matter domain era admits such symmetry for arbitrary power of Ricci scalar ($\Re$) in Palatini $F(\Re)$ gravity. It has been pointed out that their result is completely wrong, since the conserved current thus obtained does not satisfy the field equations in general. The only one Noether solution is found in the matter dominated era, which admits a coasting solution in the form $a = a_{0} t$ , which although fits SNIa data perfectly \cite{sar1}. \\

\end{document}